\documentclass{jfm}
\usepackage{graphicx}
\usepackage{epstopdf, epsfig}
\usepackage{amsmath}
\usepackage{siunitx}

\shorttitle{Viscous flow in a soft valve}
\shortauthor{K. Park et al.}

\title{Viscous flow in a soft valve}

\author{K. Park\aff{1}, A. Tixier\aff{2}, A. H. Christensen\aff{1}, S. F. Arnbjerg-Nielsen,\\ M. A. Zwieniecki\aff{2} \and K. H. Jensen\aff{1} \corresp{\email{khjensen@fysik.dtu.dk, mzwienie@ucdavis.edu}} }

\affiliation{\aff{1}Department of Physics, Technical University of Denmark, DK-2800 Kgs. Lyngby, Denmark
\aff{2}Department of Plant Sciences, University of California, Davis, CA 95616, USA}

\begin{document}

\maketitle

\begin{abstract}
Fluid-structure interactions are ubiquitous in nature and technology. However, the systems are often so complex that numerical simulations or ad hoc assumptions must be used to gain insight into the details of the complex interactions between the fluid and solid mechanics. In this paper, we present experiments and theory on viscous flow in a simple bioinspired soft valve which illustrate essential features of interactions between hydrodynamic and elastic forces at low Reynolds numbers. The setup comprises a sphere connected to a spring located inside a tapering cylindrical channel. The spring is aligned with the central axis of the channel and a pressure drop is applied across the sphere, thus forcing the liquid through the narrow gap between the sphere and the channel walls. The sphere's equilibrium position is determined by a balance between spring and hydrodynamic forces. Since the gap thickness changes with the sphere's position, the system has a pressure-dependent hydraulic resistance. This leads to a non-linear relation between applied pressure and flow rate: flow initially increases with pressure, but decreases when the pressure exceeds a certain critical value as the gap closes. To rationalize these observations, we propose a mathematical model that reduced the complexity of the flow to a two-dimensional lubrication approximation. A closed-form expression for the pressure-drop/flow rate is obtained which reveals that the flow rate $Q$ depends on the pressure drop $\Delta p$, sphere radius $a$, gap thickness $h_0$, and viscosity $\eta$  as $Q\sim \eta^{-1} a^{1/2}h_0^{5/2}\left(\Delta p_c-\Delta p\right)^{5/2}\Delta p$, where the critical pressure $\Delta p_c$ scales with the spring constant $k$ and sphere radius $a$ as $\Delta p_c\sim k a^{-2}$. These predictions compared favorably to the results of our experiments with no free parameters. 
\end{abstract}
\begin{keywords}
fluid-structure interactions, lubrication theory, low-Reynolds-number flow valve, biofluiddynamics
\end{keywords}
\section{Introduction}

Fluid flow in flexible geometries occurs in many biological, medical, and industrial environments \citep{duprat2015fluid}. Systems that couple the elasticity of a solid body with the motion of a fluid often have intersting properties; for instance, interactions between fluid-mechanical and elastic forces can lead to nonlinear pressure-drop/flow-rate relations \citep{grotberg2004biofluid,heil2015flow} and the generation of instabilities \citep{heil2011fluid}, even in relatively simple channel geometries.


Soft channels and valves play important roles in biological fluid systems: from flexible veins that carry nutrients in animals \citep{heil2011fluid} to valves which maintain unidirectional flow through the human heart \citep{sotiropoulos2016fluid}. Soft valves are also relevant to fluid transport in plants, for instance, in torus-margo pit pores, microscopic channels that permit water flow between neighboring xylem tracheid conduits in plants \citep{choat2008structure,jensen2016sap}. The pore structure prevents air embolisms from spreading inside the plant by blocking the pore with a circular disc if a large pressure difference develops between adjacent tracheids. The disc is held in place by soft cellulose fibers which can deform elastically. This soft geometry may modify flow even in the absence of air embolisms, however, previous studies have assumed either a static geometry \citep{lancashire2002modelling,schulte2012computational} or inviscid flows \citep{chapman1977hydrodynamical}, and the effect of fluid-structure interactions at low Reynolds numbers on this system is not currently well understood. 

Systems that couple the elasticity of a solid body with the motion of a fluid link two sets of dynamics that, even when treated separately, can be quite complicated. The properties of biological channels and valves have been studied extensively, however, the geometry is often so complex that numerical simulations or ad hoc assumptions must be used to gain insight into the details of the complex interactions between the fluid and solid mechanics \citep{bellhouse_talbot_1969,mcculloh2003water,kim2012natural,park2014optimal,yang2014duration,gart2015dogs,jensen2016sap}. Studies on models systems inspired by nature provide a promising alternative to \emph{in-vivo} experiments or full-scale numerical simulations. For instance, experiments on microfluidic devices have demonstrated that fluid-induced deformation of elastic sheets and fibres in channels can influence the flow at low Reynolds numbers \citep{wexler2013bending,ledesma2014experimental}, and similar results have been obtained using external mechanical actuation \citep{holmes2013control}. 

In this paper, we present  experiments and theory on a simple spring-actuated valve system inspired by biology. It captures many of the non-linear interactions that characterizes flow in soft geometries where the hydraulic resistance varies with applied pressure. The valve is suitable for regulation of small-scale fluid flows and relevant to several biofluiddynamic problems. Moreover, the simple geometry allows us to derive a closed form solution, based on lubrication theory, which is compared to experimental data. 
\begin{figure}
\begin{center}
\includegraphics[width=13cm]{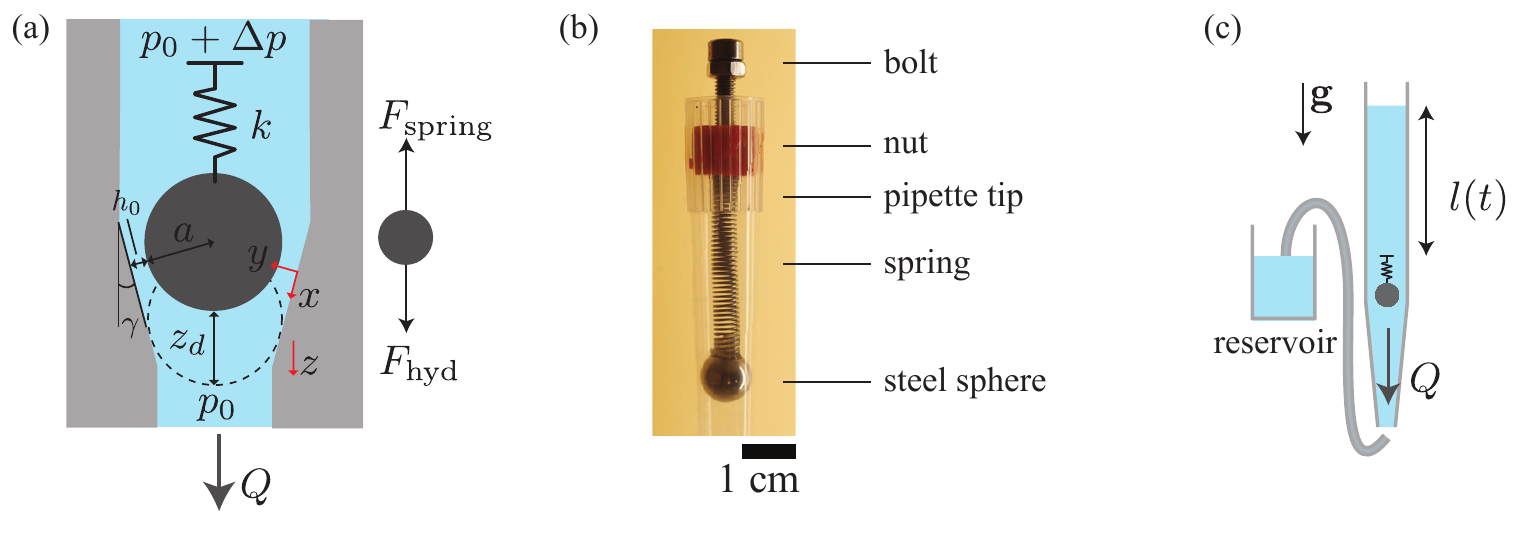}
\caption{Schematics of the soft vale system and the experimental setup. (a) Schematic of the spring-actuated soft valve. A sphere of radius $a$ is suspended from a spring (spring constant $k$) inside a tapering channel. The sphere's position is determined by force equilibrium between hydrodynamic and force and spring forces. (b) Photograph of the setup showing the steel sphere, extension spring and pipette tip. (c) Schematic of the pressure-drop/flow-rate measurement system. The pressure-drop is controlled by the liquid column height $l(t)$, and the flow rate $Q=\pi (D/2)^2 \text dl/\text dt$ and pressure drop $\Delta p = \rho g l(t)$ is determined by image analysis (see additional details in text).}
\label{Fig:Fig1}
\end{center}
\end{figure}

\section{Experiment}
\label{sec:experiments}
We consider the effect of couplings between flow and elasticity on the idealized soft valve shown in Fig. \ref{Fig:Fig1}(a):
 a sphere connected to an extension spring located inside a tapering cylindrical pipe. The spring is aligned with the central axis of the channel and a pressure drop $\Delta p$ is applied across the sphere. 
  The sphere's equilibrium position is determined by a balance between elastic and hydrodynamic forces. This leads to a non-linear relation between applied pressure $\Delta p$ and flow rate $Q=\Delta p /R(\Delta p)$, where $R(\Delta p)$ is the pressure-dependent hydraulic resistance.

\subsection{Experimental methods}
The valve was fabricated using a steel sphere of radius $a=5$ mm (KVJ Aktieselskab, 45RB5) which was glued onto an extension spring (spring constant $k = 0.16$ N/mm, RS Components 821-273), see Fig. \ref{Fig:Fig1}(b). The spring and sphere was anchored to a M4 bolt that allowed us to control the vertical position of the sphere 
along the central axis of a 3 mL pipette tip of opening angle $\gamma = 0.024$ rad (alpha laboratories, LW8930). Finally, both ends of the the pipette was connected to silicone tubing (inner diameter $D=12$ mm).

We used the hydrostatic pressure difference provided by a vertical liquid column of height $l(t)$ in the silicone tubing to drive flow across the valve, thus sweeping a broad range of pressure drops $\Delta p$ in each experiment (Fig. \ref{Fig:Fig1}(c)). 
The liquid column height $l(t)$ was tracked over time using image analysis of photographs captured using a digital camera (Canon, EOS 5D MARK III, 35mm Sigma lens  F1.4 DG HSM Art) as described in Appendix \ref{app:tracking}. The pressure drop $\Delta p = \rho g l(t)$ and flow rate $Q=\pi (D/2)^2 \text dl/\text dt$ was subsequently determined, where $\rho$ is the liquid density and $g$ is the gravitational acceleration (Fig. \ref{Fig:Fig1}(c)). In the experiments, we varied the initial ball position in the range $z_d\sim2.5-4.5$ mm (corresponding to initial gap thicknesses of $h_0 = z_d \sin \gamma \sim 60-110 \,\mu$m), and the liquid viscosity from DI water (1 cSt) to Silicon oil (100 cSt).

\subsection{Observations}
\begin{figure}
\begin{center}
\includegraphics[width=6.5cm]{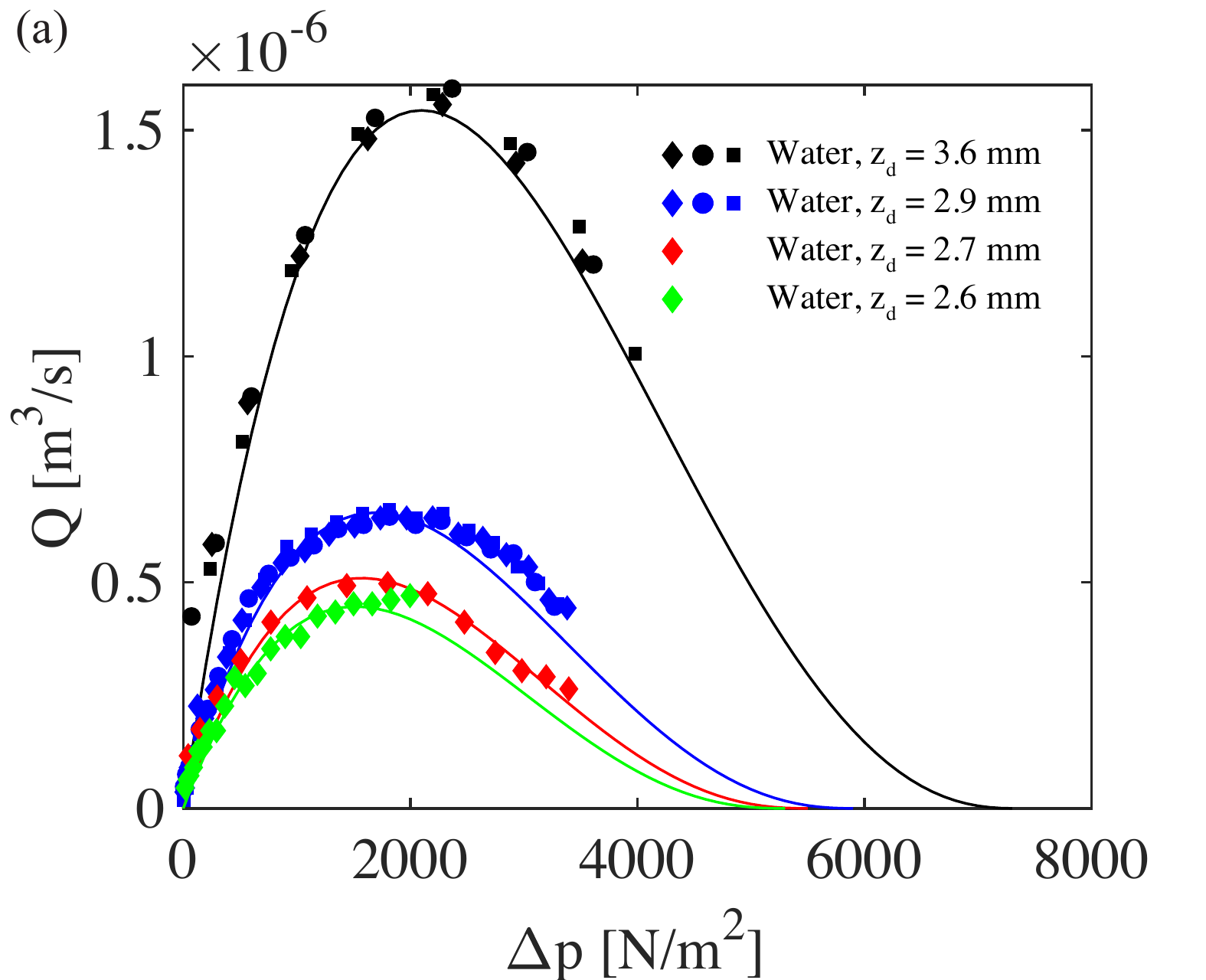}\hfill
\includegraphics[width=6.5cm]{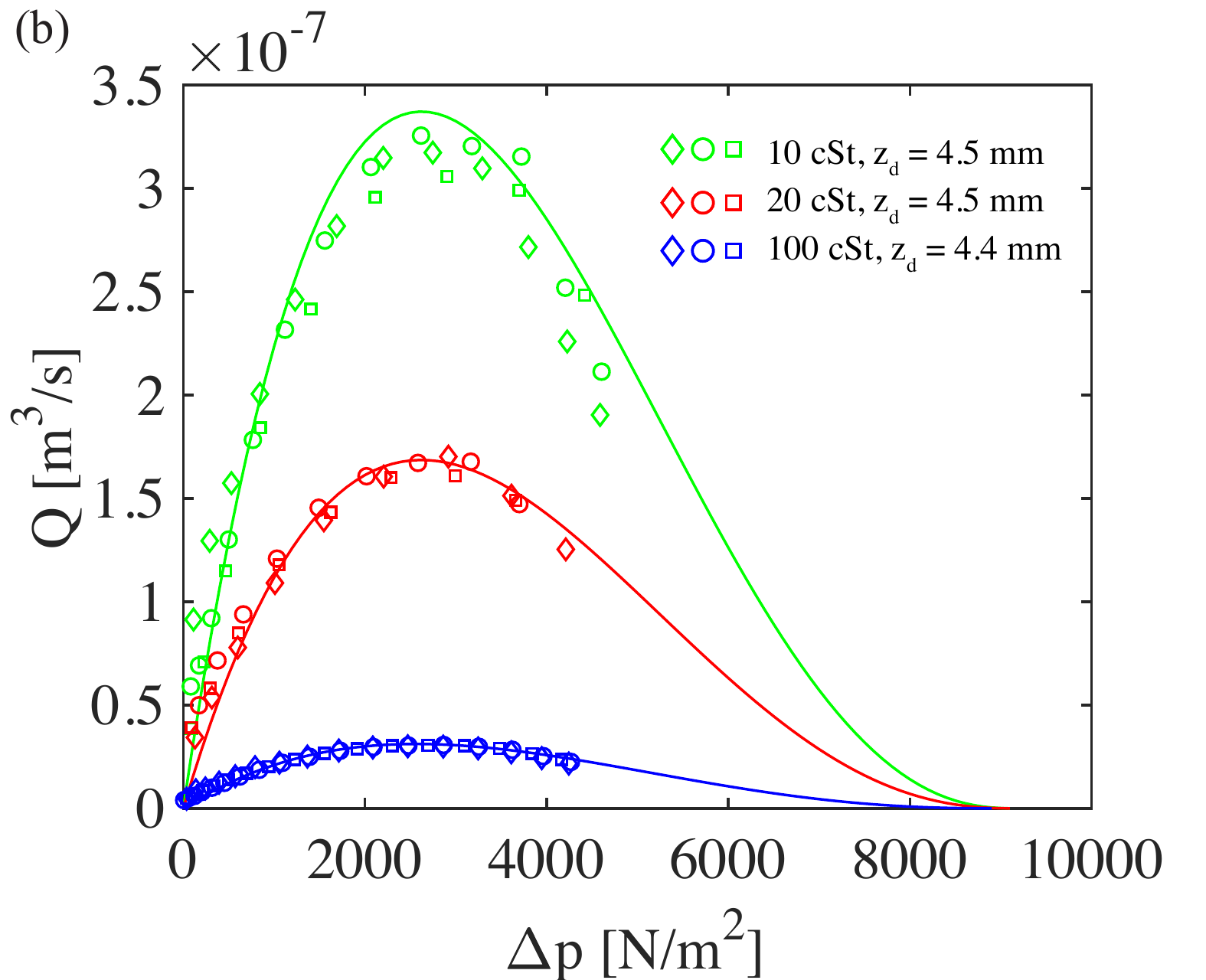}
\caption{Pressure-drop/flow rate relation for viscous flow in a soft valve. (a) Flow rate $Q$ plotted as a function of applied pressure $\Delta p$ for DI water flow with viscosity $\eta = 1 $ cSt, spring constant $k=0.16$ N/mm, and sphere radius $a=5$ mm. Results for initial ball position $z_d = 2.6, \,2.7, \,2.9,$ and$\,3.6$ mm are shown as green, red, blue and black symbols, respectively. (b) Flow rate $Q$ plotted as a function of applied pressure $\Delta p$ for silicone oil with initial ball position $z_d\sim4.5$ mm, spring constant $k=0.16$ N/mm, and sphere radius $a=5$ mm. Results for viscosities $\eta = 10,\, 20,$ and$\, 100 $ cSt are shown as green, red, and blue symbols, respectively. 
}
\label{Fig:Fig2}
\end{center}
\end{figure}

Our experiments reveal that  the flow rate increases linearly with pressure when the pressure drop is relatively small, because the position of the sphere is not influenced by the flow (Fig. \ref{Fig:Fig2}). However, as pressure increases, hydrodynamic forces push the sphere towards the channel wall, thus reducing the gap size $h_0$ (Fig. \ref{Fig:Fig1}(a)). This leads to an increase in the resistance to flow. After reaching a maximum in flow $Q_\text{max}$, at the pressure $\Delta p_\text{max}$, the flow rate starts to decrease with increasing pressure and slowly approaches $Q=0$ at the critical pressure drop $\Delta p_\text{c}$. The absolute magnitude of the flow can be tuned by varying the initial ball position $z_d$ (and hence thickness of the gap between the sphere and tapering channel walls) or by changing the viscosity $\eta$ of the fluid.  We observe that both the peak flow rate $Q_\text{max}$ and optimum pressure $\Delta p_\text{max}$ increases with $z_d$. By contrast, only the flow rate $Q$ appears to be affected significantly by variations in $\eta$.

The focus of our paper is predicting the flow rate as a function of the system parameters.  In most our experiments, the Reynolds number $Re={(\rho Q)}/{(2\pi\eta a)}$, is less than $\sim10^{ -3 }$, and hence viscous effects dominate the flow. We take advantage of this by using steady-state lubrication theory in the following analysis. However, we note that when the Reynolds number is increased to $\sim 10^3$ (corresponding to an initial vertical displacement of $z_d=5$ mm), periodic disturbances in the flow rate and sphere position can be observed (see appendix \ref{app:oscillation} and supplementary movie. 2). The observed pattern have features that resemble self-excited oscillations (see e.g. \cite{luo1996numerical}), but we shall leave its detailed consideration for future investigations.

\section{Theory}
\label{sec:theory}
To rationalize the experimental observations (Fig. \ref{Fig:Fig2}), we seek to characterize the pressure-drop/flow rate relation as a function of the system parameters
\begin{equation}
Q = \frac{\Delta p}{R}.\label{eq:R1}
\end{equation}
The pressure-dependent hydraulic resistance $R(\Delta p)$ is a function of the vertical position $z$ of the sphere, which in turn is determined by the force-balance on the sphere due to spring- and hydrodynamics forces along the $z$-axis (Fig. \ref{Fig:Fig1}(a)):
\begin{equation}
F_{\text{spring}} + F_{\text{hyd}} = 0. \label{eq:force1}
\end{equation}

We proceed in two steps to compute the pressure-dependent resistance $R$ in Eq. \eqref{eq:R1}: First, we use lubrication theory to determine the flow field in the narrow gap between the sphere and the tapering channel walls for a fixed vertical position of the sphere and pressure drop $\Delta p$. This leads to a position-dependent hydraulic resistance $R(z)$. Our analysis of the static problem follows \citet{smistrup2007magnetically}, who considered a magenetically acuated ball valve. Then, we use the force-balance equation \eqref{eq:force1} to solve for the equilibrium position $z_\text{eq}(\Delta p)$ which leads to an expression for $R(\Delta p)$.

The valve system comprises a sphere connected to a spring in a tapering cylindrical channel with opening angle $\gamma$ (Fig. \ref{Fig:Fig1}(a)). In the experiments, the sphere radius is $a=5\times 10^{-3}$ m while the typical size of the gap between the sphere and the channel walls is $h_0\leq10^{-4}$ m. We therefore consider the limit where the gap is narrow ($h_0 \ll a$), and let $\beta a$ denote the vertical displacement of the sphere above the point of contact where $h_0=0$ and $\beta =0$. The gap thickness can then be expressed in terms of $\beta$ and the sphere radius $a$, and opening angle $\gamma$ as $h_0=a\beta\sin(\gamma)$. Because the gap is narrow, the coordinate system can locally be considered as Cartesian, with $x$- and $y$-axes indicated in Fig.~\ref{Fig:Fig1}(a). Note that the $z$-axis in the diagram does not belong to this coordinate system. 

\subsection{Lubrication theory}
The flow is governed by the Navier-Stokes and continuity equations
\begin{align}
\rho\left[\dfrac{\partial \textbf{u}}{\partial t} + (\textbf{u}\cdot\nabla)\textbf{u}\right]&=-\nabla p + \eta \nabla^2\textbf{u}
\\ \nabla \cdot \textbf{u}&=0
\end{align}
where $\rho$ denotes the density, $\eta$ the viscosity, $p$ the pressure, and $\textbf{u}$ the velocity field. In the steady low-Reynolds-number limit, the Navier-Stokes and continuity equations for the flow in the gap reduces to 
\begin{subequations}
\begin{align}
\eta\left( \dfrac{\partial^2u_x}{\partial x^2}+\dfrac{\partial^2u_x}{\partial y^2}\right)&=\dfrac{\partial p}{\partial x}\label{eq:NVS1}\\
\eta\left( \dfrac{\partial^2u_y}{\partial x^2}+\dfrac{\partial^2u_y}{\partial y^2}\right)&=\dfrac{\partial p}{\partial y}\label{eq:NVS2}
\\ \dfrac{\partial u_x}{\partial x}+\dfrac{\partial u_y}{\partial y} &= 0\label{eq:CE}
\end{align}
\end{subequations}
We non-dimensionalize using the variables 
\begin{align}
X = \dfrac{x}{a}, \quad Y = \dfrac{y}{h_0}, \quad U_X = \dfrac{u_x}{q_0/h_0}, \quad U_Y = \dfrac{u_y}{q_0/a}, \quad P=\dfrac{p}{\eta q_0a/h_0^3}
\end{align} 
where $q_0$ is a constant flow rate per unit width. Equations \eqref{eq:NVS1}-\eqref{eq:CE} then become
\begin{subequations}
\begin{align}
\left(\dfrac{h_0}{a}\right)^2\dfrac{\partial^2U_X}{\partial X^2}+\dfrac{\partial^2U_X}{\partial Y^2}=\dfrac{\partial P}{\partial X}
\\\left(\dfrac{h_0}{a}\right)^4\dfrac{\partial^2U_Y}{\partial X^2}+\left(\dfrac{h_0}{a}\right)^2\dfrac{\partial^2U_Y}{\partial Y^2}=\dfrac{\partial P}{\partial Y}\\
\dfrac{\partial U_X}{\partial X}+\dfrac{\partial U_Y}{\partial Y} &= 0
\end{align}
\end{subequations}
Taking the narrow-gap-limit ($h_0/a \ll 1$), the dimensional lubrication equations are obtained
\begin{subequations}
\begin{align}
 \eta \dfrac{\partial^2u_x}{\partial y^2}&=\dfrac{\partial p}{\partial x} \label{eq:lub1} \\
0 &= \dfrac{\partial p}{\partial y}\label{eq:lub2} \\
\dfrac{\partial u_x}{\partial x}+\dfrac{\partial u_y}{\partial y} &= 0\label{eq:lub3}
\end{align}
\end{subequations}

From equation \eqref{eq:lub2} it is seen that the pressure does not depend on $y$, and hence $u_x$ can be found from equation \eqref{eq:lub1}, by using no-slip boundary conditions at $y=0$ and $y=h(x)$, as
\begin{align}
u_x(x,y) = \dfrac{1}{2\eta}\dfrac{\partial p}{\partial x}(y^2-yh(x)).\label{eq:u_x}
\end{align}
The gap $h(x)$ between the sphere and the channel walls is locally approximated by a parabola
\begin{align}
h(x) = h_0\left[1+\dfrac{x^2}{2h_0a}+\mathcal{O}\left(\dfrac{x^4}{h_0a^3}\right)\right].
\end{align}
The total flow rate, $Q$, is found by integrating the flow field (Eq. \eqref{eq:u_x}) in the gap around sphere:
\begin{align}
Q = 2\pi a \cos(\gamma) \int\limits_0^{h(x)} u_x(x,y)\,\text dy = -\dfrac{\partial p}{\partial x}\dfrac{\pi a \cos(\gamma) h^3(x)}{6\eta}\label{eq:Q2},
\end{align}
where the factor $2\pi a \cos(\gamma)$ comes from integrating along the circular gap of radius $a\cos(\gamma)$.  Since the flow rate must be independent of $x$, the pressure drop across the sphere is
\begin{subequations}
\begin{align}
\Delta p =\int_{\infty}^{-\infty}\dfrac{\partial p}{\partial x}\,\text dx &= \dfrac{6\eta Q}{\pi a \cos(\gamma)}\int_{-\infty}^{\infty} \dfrac{1}{h^3(x)} \,\text dx  \\ &=\dfrac{9\eta Q}{2\sqrt{2}\cos(\gamma)(ah_0^5)^{1/2}} \\
&=\frac{9\eta Q}{2\sqrt 2\,a^3\cos \gamma (\beta \sin \gamma)^{5/2}}\label{eq:deltap},
\end{align}
\end{subequations}
where we have used that the gap size can be written as $h_0 = \beta a \sin (\gamma)$ in the final step. 
Note that the hydraulic resistance $R=\Delta p /Q$ scales with the gap viscosity, thickness and sphere radius as $R\sim \eta a^{-1/2}h_0^{-5/2}$.
\subsection{Force balance}
To derive the flow-rate/pressure drop relation for the system, we proceed to determine the equilibrium position $z_{d,\text{eq}}$ of the sphere from the force balance Eq. \eqref{eq:force1}, see Fig. \ref{Fig:Fig1}(a). This leads to an expression for the equilibrium relative displacement $\beta_\text{eq}=z_\text{eq}/a$ which then used in Eq. \eqref{eq:deltap} to determine the resistance.

To model the extension spring we assume a linear spring force of
\begin{equation}
F_{\text{spring}} = -k\Delta z,\label{eq:spring2}
\end{equation}
where $k$ is the spring constant and $\Delta z$ is the vertical displacement from the equilibrium position. Let $z_0$ be the equilibrium length when no pressure is applied and let $z_\text{max}$ be the maximum extension of the spring, obtained when the sphere touches the inclined wall. For a sphere located a distance $z_d=\beta a$ above the contact point, the spring extension is $\Delta z = z_\text{max}-z_0-\beta a$ and hence the spring force is given by $F_{\text{spring}} = -k(z_\text{max}-z_0-\beta a)$.



The hydrodynamic force on the sphere is given by 
\begin{align}
F_{\text{hyd},i}=\int_{\partial \Omega}\left[-p\delta_{ij}+\eta(\partial_iu_j+\partial_ju_i)\right] n_j\,\text da\label{eq:F_general}
\end{align}
where we use the index notation and $\partial \Omega$ denotes the surface of the sphere. In the lubrication limit, we expect that the pressure-term dominates the vertical force on the sphere. The main pressure gradient occurs in a narrow region near the gap, and the pressure-force can thus be approximated by assuming constant pressures above ($p_0+\Delta p$) and below  ($p_0$) the sphere:


\begin{align}
F_{\text{hyd}}^{\text{(p)}}&\simeq  2\pi a^2(p_0 + \Delta p) \int_0^{\pi/2+\gamma}\cos(\theta)\sin(\theta)d\theta + 2\pi a^2 p_0 \int_{\pi/2+\gamma}^\pi \cos(\theta)\sin(\theta)d\theta \nonumber
\\&=\pi a^2\cos(\gamma)^2\Delta p \label{eq:F_p_simple}
\end{align}
where $\theta$ is the angle measured from the vertical. To obtain the viscous force, we note that in the lubrication approximation $\partial/\partial x \ll \partial/\partial y$, so the dominant term is
\begin{align}
F_{\text{hyd}}^{\text{(visc)}} &\simeq 2\pi a \cos(\gamma)\eta \int_{-\infty}^\infty\left.\dfrac{\partial u_x}{\partial y}\right|_{y=h(x)}\cos(\gamma)\,\text dx = \dfrac{4\pi }{3}  \cos^2(\gamma)\sin(\gamma) \beta a^2 \Delta p 
\label{eq:F_v}
\end{align}
where the prefactor $2\pi a \cos \gamma$ comes from integration around the contact line, and the factor $\cos \gamma$ in the integral from the force projection along the $z$-axis.
Comparing the magnitude of the pressure (Eq. \eqref{eq:F_p_simple}) and viscous (Eq. \eqref{eq:F_v}) forces, we find
\begin{equation}
\frac{F_{\text{hyd}}^{\text{(visc)}} }{F_{\text{hyd}}^{\text{(p)}}} = \frac{4}{3}\frac{h_0}{a}.
\end{equation}
In our case, where $h_0\ll a$, the pressure force is seen to dominate the hydrodynamic force on the sphere. 


To find the vertical equilibrium position of the sphere, we now combine the spring force \eqref{eq:spring2} and the pressure force \eqref{eq:F_p_simple} using the force-balance  equation Eq. \eqref{eq:force1}, and find
\begin{align}
\pi a^2\cos(\gamma)^2\Delta p 
-k(z_d-a\beta)&=0
\end{align}
where we have written the initial height as $z_d = z_0-z_{\text{max}}$. This leads to an expression for the relative position $\beta$ at equilibrium:
\begin{equation}
\beta = \frac{kz_d-\Delta p \pi a^2 \cos (\gamma)^2}{ka} \label{eq:beta1}.
\end{equation}
Combining Eqns. \eqref{eq:beta1} and \eqref{eq:deltap} leads to a non-linear pressure-dependent flow rate $Q$
\begin{equation}
Q = \dfrac{2\sqrt{2}}{9}\dfrac{\cos^6(\gamma)a^3}{\eta}\left(\dfrac{a\pi\sin(\gamma)}{k}\right)^{5/2} \left(\dfrac{kz_d}{\pi a^2 \cos^2(\gamma)}-\Delta p\right)^{5/2}\Delta p.\label{eq:masterQ}
\end{equation}
Introducing the low-pressure-limit resistance 
\begin{equation}
R_0 = \frac{9}{2\sqrt 2 \sin (\gamma)^{5/2}\cos(\gamma)}\frac{\eta }{(az_d^5)^{1/2}}
\end{equation}
and the critical pressure drop $\Delta p_c$ that makes the ball block the channel
\begin{equation}
\Delta p_c = \dfrac{kz_d}{\pi a^2 \cos^2(\gamma)}\label{eq:p_max},
\end{equation}
leads to a convenient expression for the flow rate:
\begin{equation}
Q = \frac{\Delta p}{R_0}\left(1-\Delta p/\Delta p_c\right)^{5/2}.\label{eq:sol}
\end{equation}
When $\Delta p \ll \Delta p_c$ the flow rate scales linearly with pressure drop $\Delta p$, in accord with experimental observations (Fig. \ref{Fig:Fig2}). A maximum in the flow rate occurs at the pressure drop $\Delta p_{\text{max}}=\frac 27 \Delta p _c$, above which the valve gradually closes and the flow rate $Q$ starts to decay.

\begin{figure}
\begin{center}
\includegraphics[height=6cm]{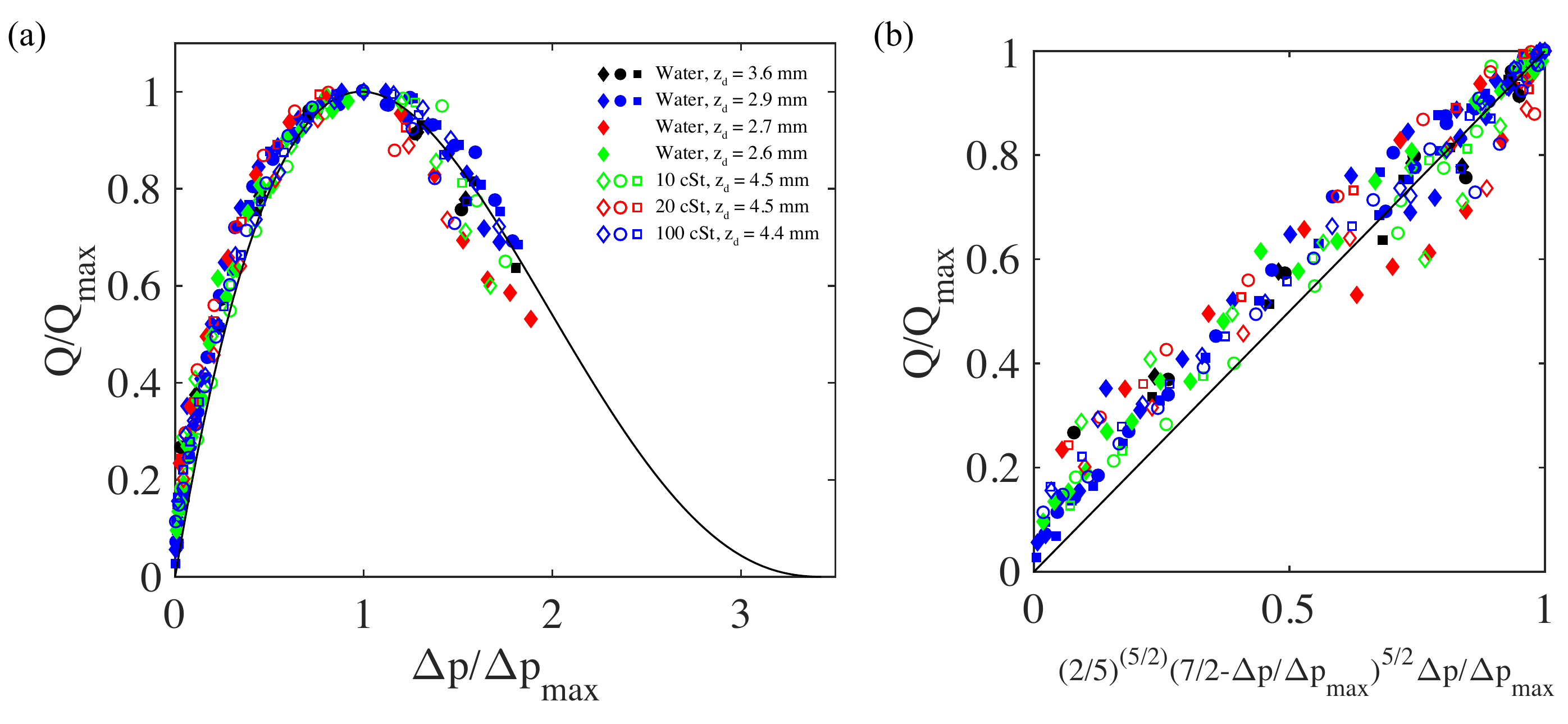}
\caption{Normalized flow rate $Q/Q_{\text{max}}$ plotted as a function of (a) relative pressure drop $\Delta p/\Delta p_{\text{max}}$ and (b) $\left( \frac { 2 }{ 5 }  \right) ^{ 5/2 }\left( \frac { 7 }{ 2 } -\frac { \Delta p }{ \Delta p_{ max } }  \right) ^{ 5/2 }\frac { \Delta p }{ \Delta p_{ max } }$,  using data from Fig. \ref{Fig:Fig2} (a,b). The solid line shows the theoretical prediction obtained using lubrication theory (Eq. \eqref{eq:rescaledEQ}).}
\label{Fig:Fig4}
\end{center}
\end{figure}

\subsection{Comparison between experiment and theory}
 
The lubrication-theory model for the flow rate $Q$ (Eq. \eqref{eq:masterQ}) captures the essential qualitative behavior of the experiment data (Fig. \ref{Fig:Fig2}), and we find good quantitative agreement between experiments and theory with no free parameters. Moreover, the data collapsed onto a single line when scaled according to the maximum flow rate $Q_\text{max}$ and corresponding pressure $\Delta p_{\text{max}}$ (Fig. \ref{Fig:Fig4}). The data collapse is seen to be in good accord with the prediction from Eq.~\eqref{eq:sol}, which yields
\begin{equation}
\frac{Q}{Q_{\text{max}}} 
=\left(\frac{\Delta p_c -\Delta p}{\Delta p_c -\Delta p_\text{max}}\right)^{5/2}\frac{\Delta p}{\Delta p _\text{max}}
=\left(\frac{2}{5}\right)^{5/2}\left(\frac 72-\frac{\Delta p}{\Delta p _\text{max}}\right)^{5/2}\frac{\Delta p}{\Delta p _\text{max}} \label{eq:rescaledEQ},
\end{equation}
where 
we have used that the maximum pressure is $\Delta p _\text{max} = 2/7 \Delta p_c$.
\label{sec:expvstheory}

\section{Discussion and conclusion}
In this paper we have presented an experimental and theoretical study of a bioinspired soft valve. The model system comprised a sphere connected to a spring located inside a tapering channel. A coupling between elasticity and fluid motion determines the hydraulic resistance of the valve because the position of the sphere is influenced by the flow and vice versa. Using a simple setup, we performed experiments to characterize the pressure-drop/flow rate relationship for the system. Our data revealed a strongly non-linear behaviour and the existence of an optimal pressure drop for the system where the flow rate is maximized.

Assuming low-Reynolds-number flow, we proposed a mathematical model that reduced the complexity of the flow to a two-dimensional lubrication approximation. The model revealed a relatively complete picture of the geometric and material parameters that affect flow through the system. The model predictions compared favorably to the results of our experiments. Our analysis is limited to steady viscous flows, though preliminary observations suggest onset of self-excited oscillations at higher Reynolds numbers.


\section*{Acknowledgements}
\noindent This work was supported by a research grant (13166) from VILLUM FONDEN.


\bibliographystyle{jfm}
\bibliography{2017_BallValveJFM}

\newpage
\appendix

\section{Tracing the meniscus}
\label{app:tracking}
\begin{figure}
\begin{center}
\includegraphics[height=5cm]{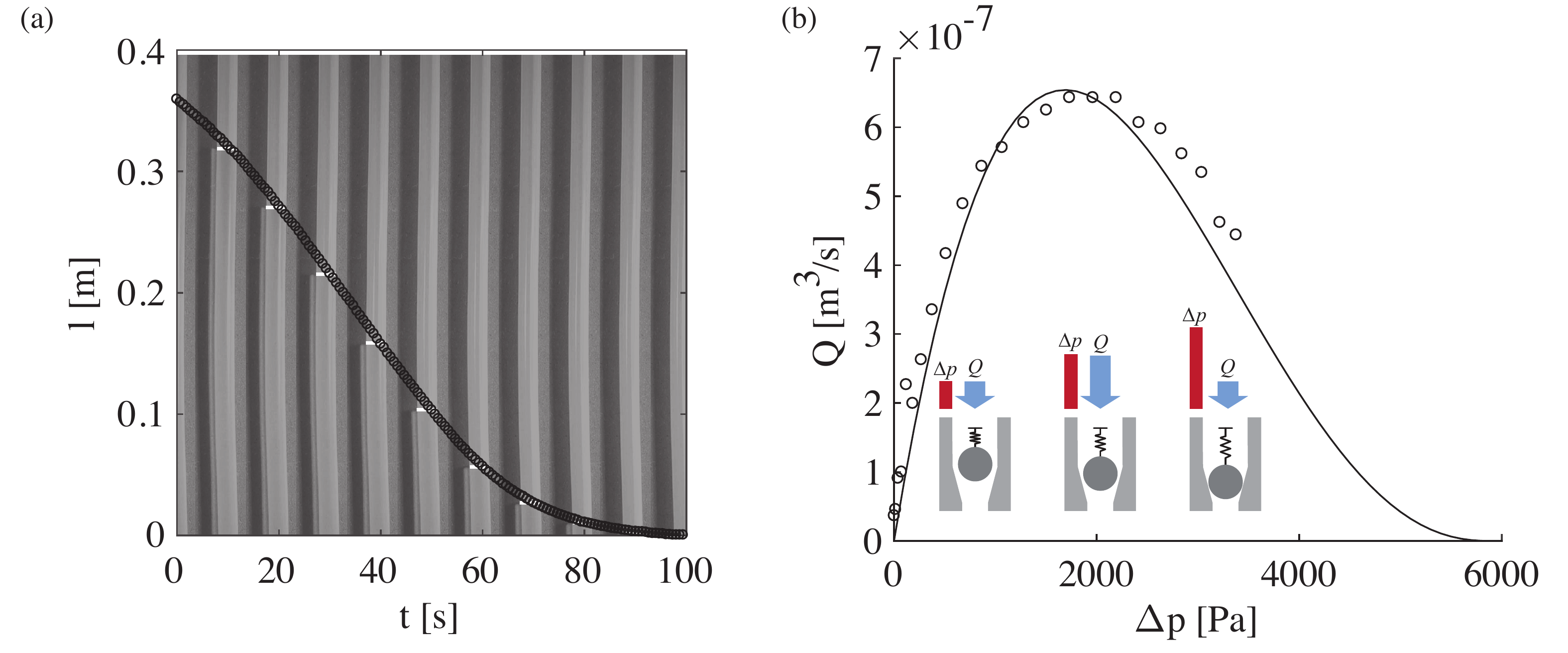}
\caption{Experimental result of $z_d$ = 2.9 mm with water. (a) Time vs liquid column height graph with a montage of every 10s images. Tracking points (Circles) is well fitted with experimental images (Horizontal white lines). (b) Pressure drop and flow rate graph from (a). inset shows the ball position at certain pressure drop.}
\label{fig:window}
\end{center}
\end{figure}

The pressure-drop/flow rate relation was determined experimentally as outline in Fig.\ref{Fig:Fig1}(c) and Fig. \ref{fig:window}. A stack of images showing the meniscus movement was converted to gray-scale and cropped to show the silicone tube above the soft valve. The 2D-images $I(z,x,t)$ were converted to a 1D intensity profile $I'(z,t)=\sum_x I(z,x,t)$ by summing the values in each horizontal image row.  We traced the position of the meniscus by locating the maximum in the intensity gradient $\sim \partial_z I'(z,t)$ at the position $z_{\text{max}}(t)$. This yielded the meniscus position $z_{\text{max}}=l(t)$ which in turn allowed us to determine the flow rate $Q$. In order to minimize effects of image noise in determining $z_{\text{max}}$, we manually indicated the initial meniscus position and used a tracing window. 

\begin{figure}
\label{app:oscillation}
\begin{center}
\includegraphics[width=11cm]{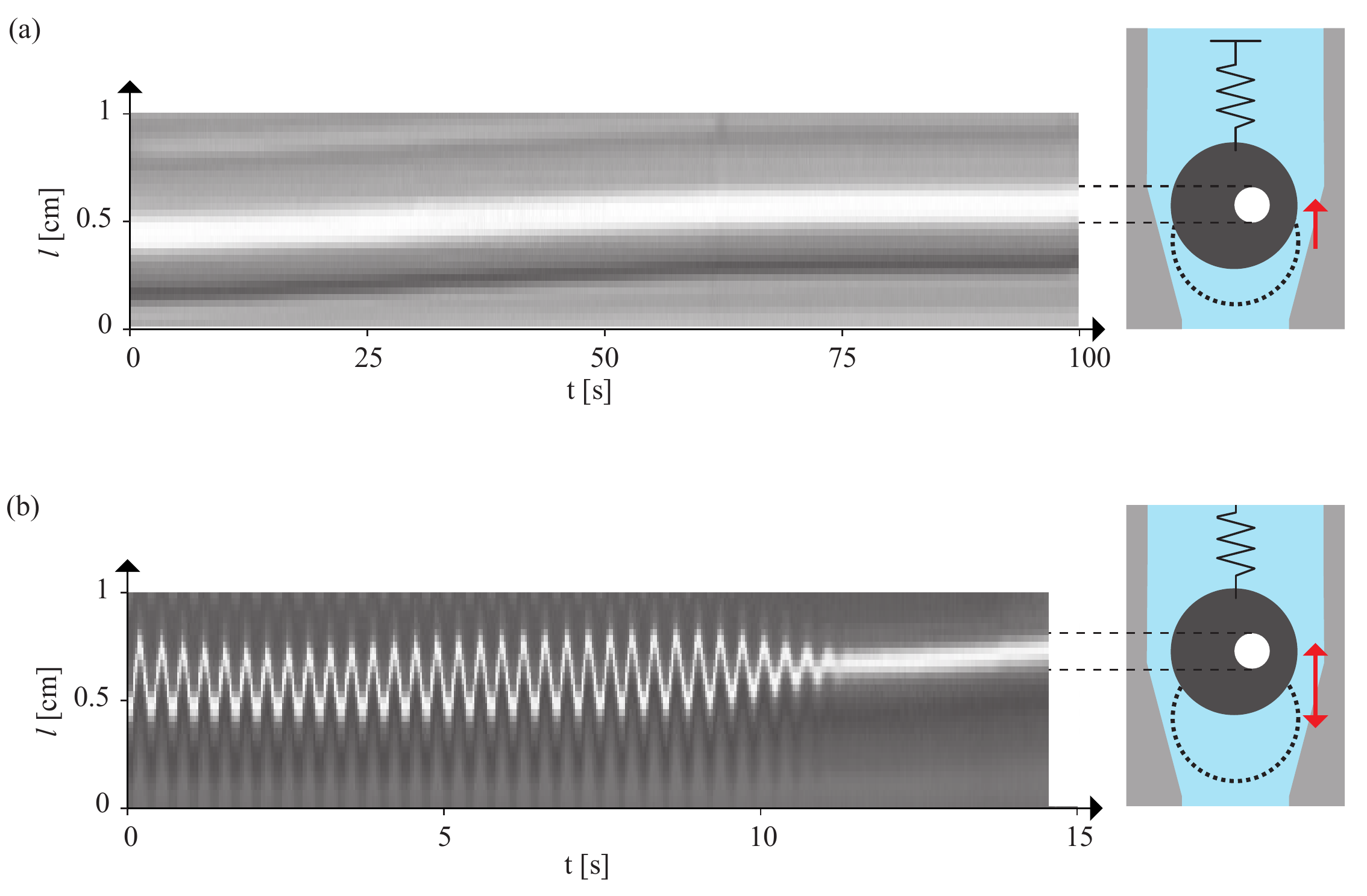}
\caption{(a) Experimental kymograph of a soft valve with water and $z_d$ = 2.9 mm. Image variation shows ball movement during the experiment (see Supplemental movie1). (b) Kymograph of a soft valve with water and $z_d$ = 5.4 mm. During the first 10 seconds the valve oscillates and then moves softly. Pressure drop decreases from 7500 Pa. The oscillation stops when the Reynolds number between the gap is less than 2700 (see Supplemental movie2). }

\end{center}
\end{figure}

\end{document}